\documentclass[12pt]{article} 
\usepackage[a4paper, margin=1in]{geometry} 
\usepackage{amsmath,amssymb,amsfonts} 
\usepackage{algorithmic} 
\usepackage{graphicx} 
\usepackage{color} 
\usepackage{textcomp} 
\usepackage{xcolor} 
\usepackage{algorithm} 

\usepackage[caption=false]{subfig}
\usepackage{cleveref}
\usepackage{multirow}  
\usepackage{array}
\usepackage[export]{adjustbox}

\def\BibTeX{{\rm B\kern-.05em{\sc i\kern-.025em b}\kern-.08em
    T\kern-.1667em\lower.7ex\hbox{E}\kern-.125emX}}
\AtBeginDocument{\definecolor{tmlcncolor}{cmyk}{0.93,0.59,0.15,0.02}\definecolor{NavyBlue}{RGB}{0,86,125}}
\newcommand{\red}[1]{{\color{red} #1}}
\newcommand{\blue}[1]{{\color{blue} #1}}

\DeclareMathOperator*{\argmin}{argmin}
\newcommand{\etal}{\textit{et al.}}
\newcommand{\ie}{\textit{i.e.}}
\newcommand{\eg}{\textit{e.g.}}

\begin{document}

\title{SDAT: Sub-Dataset Alternation Training for Improved Image Demosaicing}

\author{Yuval Becker\\
\and
Raz Z. Nossek\\
\and
Tomer Peleg\\
\and 
Samsung\\
Samsung Israel R\&D Center, Tel Aviv, Israel\\
}



\maketitle  
\begin{abstract}
Image demosaicing is an important step in the image processing pipeline for digital cameras.
In data centric approaches, such as deep learning, the distribution of the dataset used for training can impose a bias on the networks' outcome. For example, in natural images most patches are smooth, and high-content patches are much rarer. This can lead to a bias in the performance of demosaicing algorithms. 
Most deep learning approaches address this challenge by utilizing specific losses or designing special network architectures. We propose a novel approach \textbf{SDAT}, Sub-Dataset Alternation Training, that tackles the problem from a training protocol perspective. SDAT is comprised of two essential phases. In the initial phase, we employ a method to create sub-datasets from the entire dataset, each inducing a distinct bias. The subsequent phase involves an alternating training process, which uses the derived sub-datasets in addition to training also on the entire dataset.
SDAT can be applied regardless of the chosen architecture as demonstrated by various experiments we conducted for the demosaicing task. The experiments are performed across a range of architecture sizes and types, namely CNNs and transformers. We show improved performance in all cases. We are also able to achieve state-of-the-art results on three highly popular image demosaicing benchmarks. 
\end{abstract}

\section{INTRODUCTION}
\label{sec:intro}

\begin{figure}
    \centering

   \includegraphics[width=0.7\linewidth]{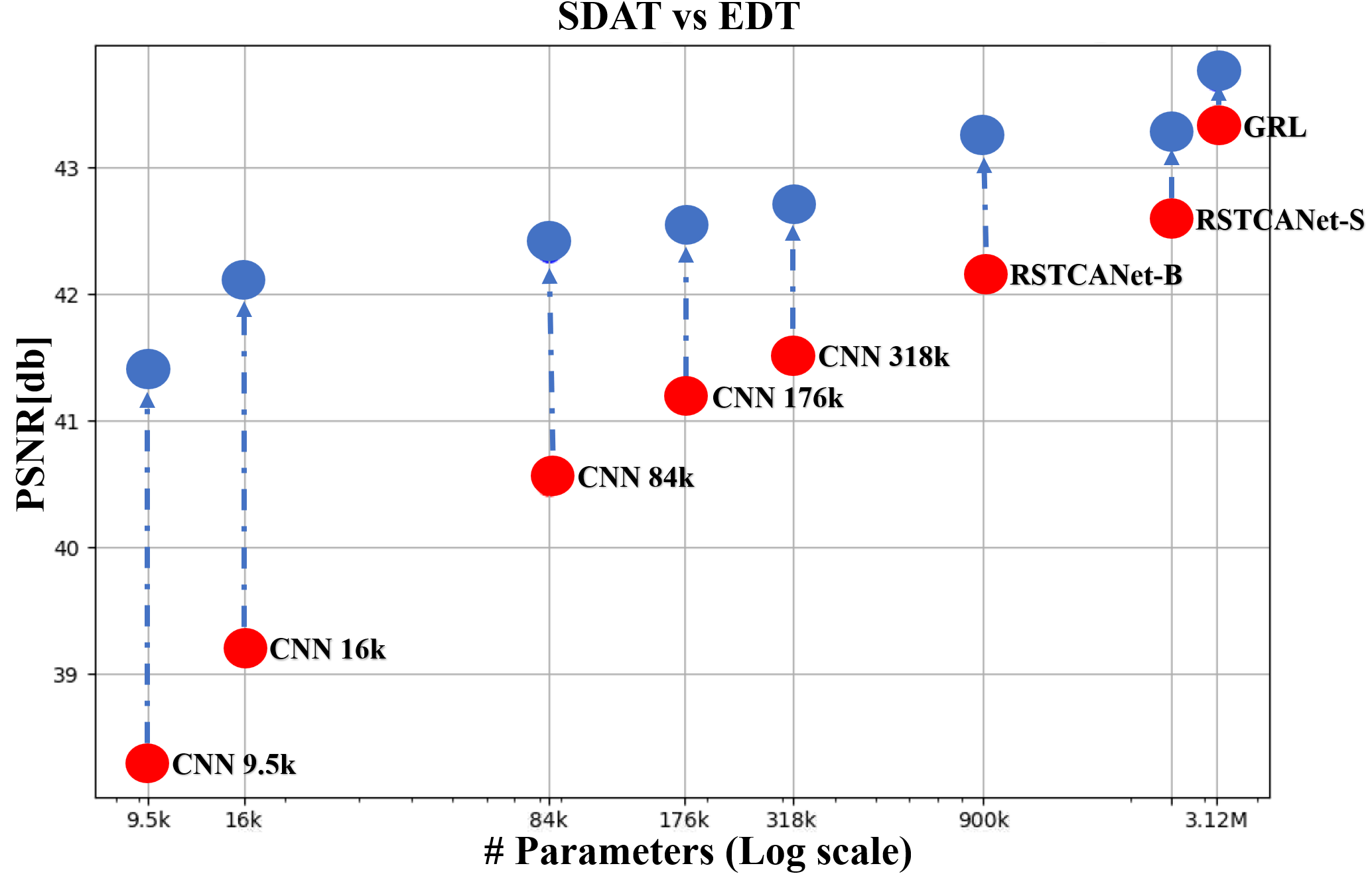}
   \caption{A comparison of PSNR results between our training method (SDAT, \blue{blue} circles) and a standard training on the entire dataset (EDT, \red{red} circles) applied on in-house and various popular architectures \cite{ xing2022residual, li2023efficient} over Kodak \cite{li2008image} dataset. By using our training method we achieved better results compared to standard training across all architectures. In addition, we achieved state-of-the-art (SOTA) results using GRL architecture.} 
   \label{fig:intro}
\end{figure}

\begin{figure*}[!ht]
\noindent
    \begin{minipage}{0.24\textwidth}
    \captionsetup{justification=centering}
    \centering
    
     \subfloat[GT image 26 from Urban100]{
                    \includegraphics[width=0.9\textwidth,valign=b]{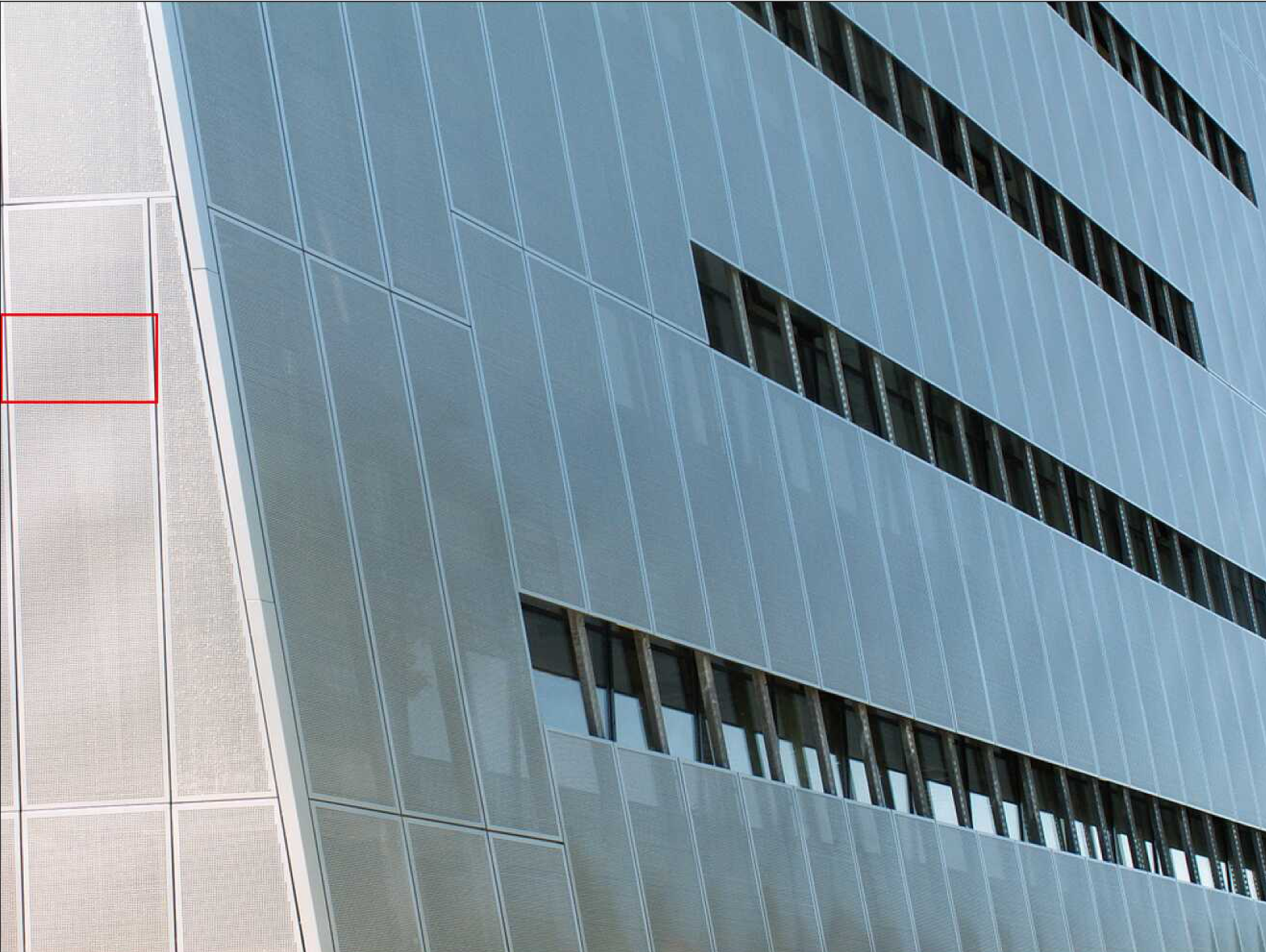}
                    \label{fig:urban26_GT}
                }
                
     
    \vspace{2mm}

    \subfloat[Crop from GT]{
                    \includegraphics[width=0.9\textwidth,valign=b]{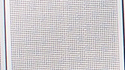}
                    \label{fig:crop_urban26_GT}
                }
                
    
    \end{minipage}%
    \hspace{2mm}
    \begin{minipage}{0.74\textwidth}
    \vspace{2mm}
        \begin{minipage}{1\textwidth}
            \captionsetup{justification=centering}
                            \label{fig:urban_gt_crop}
                \subfloat[RSTCANet-B]{
                    \includegraphics[width=0.3\textwidth,valign=b]{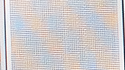}
                }
                \label{fig:urban_B}
                \subfloat[RNAN]{
                    \includegraphics[width=0.3\textwidth,valign=b]{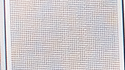}
                }
                \label{fig:tgv_ef_ipm_first_inst}
                \subfloat[RSTCANet-L]{
                \includegraphics[width=0.3\textwidth,valign=b]{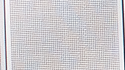}
            }
            \label{fig:RSTC_L}
        \end{minipage}%
        \vspace{5mm}
        \begin{minipage}{1\textwidth}
            
            \captionsetup{justification=centering}
            \subfloat[RSTCANet-B-SDAT]{
                \includegraphics[width=0.3\textwidth,valign=b]{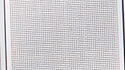}
            }
                            \label{fig:Our_RSTC_b}
            \captionsetup{justification=centering}
            \subfloat[GRL]{
                \includegraphics[width=0.3\textwidth,valign=b]{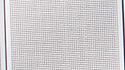}
            }
                \label{fig:GRL_orig}
            \captionsetup{justification=centering}
            \subfloat[GRL-SDAT]{
                \includegraphics[width=0.3\textwidth,valign=b]{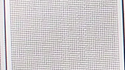}
            }
        \end{minipage}%
        \end{minipage}
			
			\captionsetup{justification=justified}
			\caption{Qualitative comparison of our method compared to other top methods: RNAN, RSTCANet, and GRL. The RNAN model has 9M parameters, RSTCANet presents 3 different model sizes: B, S, and L, having 0.9M, 3.1M, and 7.1M parameters respectively and GRL consists of 3.1M parameters. We demonstrate the results of our suggested training scheme to train the RSTCANet-B and GRL models. The RSTCANet-B-SDAT model produces superior qualitative results compared to all original RSTCANet variants while having the least amount of parameters.}
			\label{fig:qual_res}
\end{figure*}

The task of image demosaicing is a crucial step in digital photography and refers to the process of reconstructing a full-resolution RGB color image from incomplete data obtained from the use of a color filter array (CFA), such as the Bayer pattern of GRBG. In digital cameras, CFA samples only a fraction of the image information, which makes the task of demosaicing complex and challenging. 
This task is part of a larger group of image restoration problems, such as denoising, deblurring, single image super-resolution, in-painting, removing JPEG compression, etc. These tasks aim at recovering a high-quality image from degraded input data.  
Image restoration problems are usually considered ill-posed, in the sense that it is usually not possible to determine a unique solution that can accurately reconstruct the original image. This is particularly true for image demosaicing due to the fact that the red, green, and blue color channels are sampled at different locations and rates, which can cause severe aliasing issues. In recent years, the use of convolutional neural networks (CNNs) and transformers has shown significant success in addressing image restoration problems \cite{liang2021swinir, zhang2019residual,mei2023pyramid,xing2022residual, li2023efficient} in general, and for image demosaicing in particular, \cite{gharbi2016deep, kokkinos2018deep, liu2020joint, niu2018low, qian2022rethinking, tan2018deepdemosaicking, xing2022residual}. \\ It is well known that inductive bias, which refers to the assumptions and prior knowledge that the model brings to the learning process, is necessary for the model convergence \cite{mitchell1980need}. But, in some cases, it can have the opposite effect and negatively impact the model's ability to generalize. This is true for deep neural networks as well and has triggered an extensive line of study, e.g. \cite{geirhos2018imagenet, ritter2017cognitive}, just to mention a few. The bias can arise from various factors such as the network architecture, the training dataset, and others.
Our focus in this study is on the inductive bias of a given model caused by the trained dataset.
In the case of image demosaicing, biased solutions can cause artifacts, such as zippers, Moiré, color distortion, and more.
The most common cause of inductive bias is the well observed phenomenon that natural images are mostly smooth.\cite{hyvarinen2009natural, zontak2011internal}. However, there are more elements in the data that can cause a given model to induce bias which are less obvious and are more difficult to characterize.

In this paper we propose \textbf{SDAT}, Sub-Dataset Alternation Training, a novel training approach designed to guide convergence towards less biased solutions. Our method extracts sub-datasets from the entire training data, each inducing a unique bias over the trained model. Our training utilizes these sub-datasets to steer the learning process in a non-traditional manner, and prevents converging towards highly likely biased solutions. This approach involves an alternation between training on the entire dataset and training on the extracted sub-datasets.

We demonstrate how SDAT is able to improve the performance of different architectures across all evaluated benchmarks in comparison to the originally implemented training, as can be seen in \Cref{fig:intro}. One type of evaluation is following a recent trend of low-capacity models (less than 200k parameters) for edge devices performing image demosaicing \cite{ma2022searching, ramakrishnan2019deep, wang2021compact}. We show that our method is able to effectively utilize the model's capacity and surpass recent relevant works across all benchmarks using fewer number of parameters, showcasing a more efficient solution for low-capacity models. Another is demonstrating improved performance over high capacity models (in an order of 1M parameters and higher). We evaluate our method both on CNN based architectures and transformers. Thus we exemplify the effectiveness of the proposed method regardless of the architecture. Using SDAT we achieved state-of-the-art on three popular benchmarks by training a variant of the GRL architecture \cite{li2023efficient}.
 \\
 
We summarize our main contributions as follows: 
\begin{enumerate} 
    \item We introduce a novel training method for image demosaicing that is able to explore the parameter space more effectively than standard training methods, thus decreasing bias induced by the data. 
    \item We evaluate our training scheme over various model sizes and different types of architectures, showing great improvement all around and achieving SOTA results over three highly popular benchmarks for image demosacing. 
\end{enumerate}

A comparison of visual results can be found in \Cref{fig:qual_res}. 

\section{Related work}
\label{sec:RW}

 \begin{figure*}[!ht]
    \begin{center}
        \includegraphics[width=1.0\linewidth]{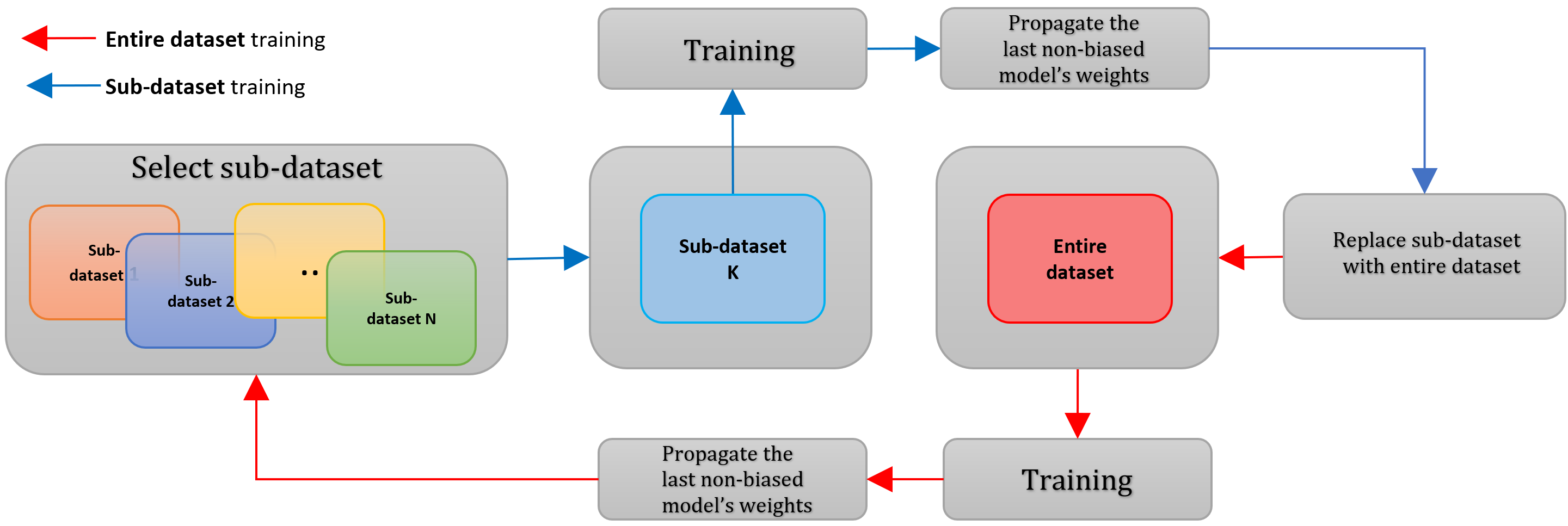}
    \end{center}    
        \caption{ An illustration of the SDAT method. Each cycle consists of two training phases. The first consists of training over a specific sub-dataset obtained from the pool of collected sub-datasets as explained in \cref{sub_sec:idn_sub_cat2}, while the second, consists of training over the entire dataset.
        Each phase is initialized by the model's weights that achieved the lowest validation loss across all sub-datasets in the previous phase.
        Every cycle a different sub-dataset is selected. The number of cycles depends on the number of categories and architecture. }
	\label{fig:optimization}   
\end{figure*}

Image demosacing aims to restore missing information from the under-sampled CFA of the camera sensor, as in most cases each pixel only captures a single color. This is a well studied problem with numerous methods suggested for solving it. The majority of methods focus on the well known Bayer pattern, which captures only one of the following colors: red, green, or blue at each pixel.\\
 Deep learning methods have become ever more popular, and there has been an active line of study dedicated to performing image demosacing alone, e.g. \cite{syu2018learning, tan2017color, xing2022residual, kerepecky2023nerd, lee2023efficient}.     
In other cases, it is joined with other tasks (such as image denoising), e.g. \cite{liu2020joint, qian2022rethinking, xing2021end, zhang2022deep}. Most of the joint methods still train a network and analyze its performance for the image demosaicing task alone. In \cite{gharbi2016deep} the authors designed a joint demosaicing and denoising network using CNNs and constructed a dataset containing solely challenging patches they mined from online data, based on demosaic artifacts. They then went on to train and test the network on cases without the presence of noise. Most works did not follow this practice and trained the demosaicing network on a general dataset (\ie without applying any mining).\\
 \textbf{Dedicated task architecture:} A prominent line of work focused on architectural modification according to the prior knowledge of the task.
Tan et al. \cite{tan2017color} proposed a CNN model that first uses bi-linear interpolation for generating the initial image and then throws away the input mosaic image. Given the initial image as the input, the model has two stages for demosaicing. The first stage estimates green and red/blue channels separately while the second stage estimates three channels jointly. The idea of reconstructing the green channel separately or reconstructing it first to use later on as a guidance map is also used by several others, such as  Cui et al. \cite{cui2018color}, and Liu et al. \cite{liu2020joint}. Other works compared the effect of convolution kernels of different sizes on the reconstruction, e.g. the work by Syu et al. \cite{syu2018learning}, which concluded the larger the size of the convolution kernel, the higher the reconstruction accuracy. Inspired by \cite{syu2018learning}, Gou et al. \cite{guo2020residual}, suggested adapting the Inception block of \cite{szegedy2017inception} to reduce computation while still having a large receptive field.\\
\textbf{General architectural modification:}
 Another recent popular line of study is utilizing a general neural network architecture and training it separately on various image restoration tasks (i.e. each task requires a separate model with different weights), such as \cite{mei2023pyramid,liang2021swinir, mou2021dynamic, zhang2021plug, zhang2019residual}. Zhang et al. suggested \cite{zhang2019residual} the RNAN architecture, which is a CNN network. Xing and Egiazarian \cite{xing2022residual} suggested slight modifications to the SwinIR architecture of \cite{liang2021swinir} for the specific task of demosaicing. PANet \cite{mei2023pyramid} performed a multi-scale pyramid attention mechanism. 
 Li et al. \cite{li2023efficient} proposed the GRL transformer-based  architecture
 for various image-to-image restoration tasks. Specifically for the demosaicing task, their suggested method achieved SOTA accuracy over KODAK \cite{li2008image}, McMaster \cite{zhang2011color}, Urban \cite{huang2015single} and CBSD68 \cite{937655} datasets. Their solution is focused on artifacts removal, as their network receives an initial demosaic solution.\\
 \begin{figure*}[!ht]
    \begin{center}
        \includegraphics[width=0.92\linewidth]{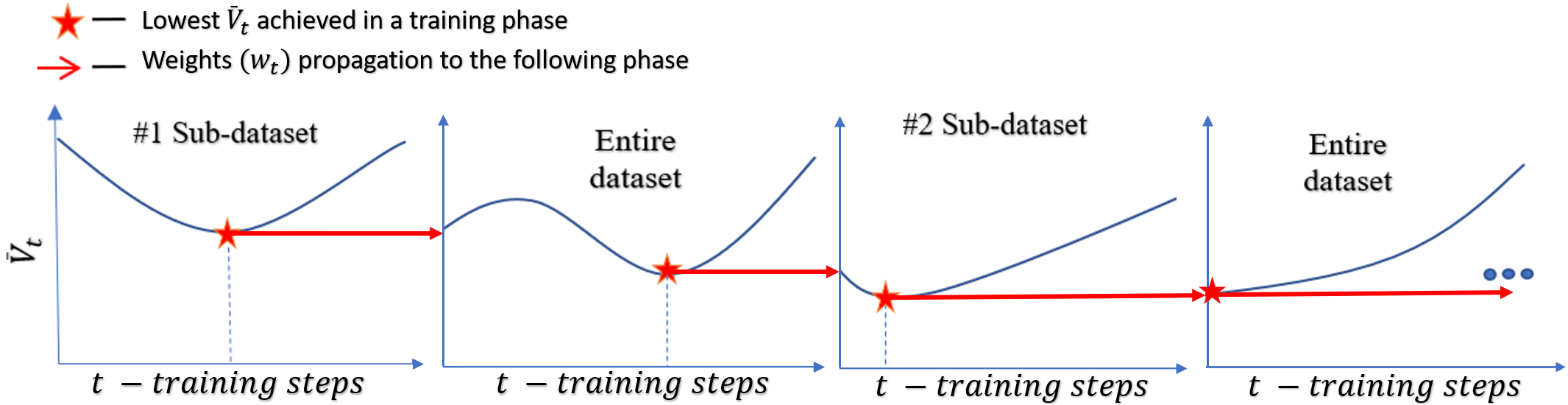}
    \end{center}
    \vspace{-2mm}
        \caption{ Depiction of the weight propagation process of SDAT. An illustration of the alternation process between a sub-dataset and the entire dataset. Each graph is a training phase over a single dataset, where the horizontal axis is the training steps and the vertical axis is the average validation loss across all sub-datasets marked with $\Bar{V}_t$.
        The star on each of the graphs marks the iteration index each training phase stopped accumulating gradients for a specific dataset, as it propagates model weights achieved at that index to the following training phase.
                Furthermore, there can be training phases that do not aggregate any gradients. As can be seen on the rightmost graph the model could not achieve a convergence that lowers $\Bar{V}_t$, therefore, it accumulated zero training steps.}
	\label{fig:sub_cat_train}   
\end{figure*}
\hspace{-1mm}\textbf{Low-capacity demosaic solutions:} 
It is noteworthy to mention another line of work that focuses on the implementation of low-capacity networks for image demosaicing. In \cite{ramakrishnan2019deep} Ramakrishnan \etal employed Neural Architecture Search (NAS) and optimization techniques, while in  \cite{wang2021compact} Wang \etal proposed a custom UNet architecture. Both approaches evaluate their solutions across different model sizes.\\
Our method (see \cref{fig:optimization} for an overview) exhibits several significant distinctions from prior research, as most of the works focused on custom losses and 
architectural modifications \cite{zhang2019residual,xing2022residual,lee2023efficient,liang2021swinir}. This work focuses on convergence and training perspectives. Unlike mining-based methods \cite{gharbi2016deep} that rely on error criteria to identify only hard examples, our approach employs different metrics and uses a novel identification mechanism to identify multiple sub-datasets, that exhibit different characteristics, each inducing a different bias into the trained model. Furthermore, our training method utilizes the sub-datasets to steer and guide convergence while training on the entire dataset.



\vspace{-1.5mm}
\section{Method} 
\label{sec:method} 

In the following, we present the required stages to perform SDAT.  
We will show that by applying our training regime we are able to better utilize the given dataset, \ie achieve better results for a given network architecture without modification to it or the loss function. 
The method involves training alternatively between the entire dataset and the various sub-datasets in a cyclic fashion. Each of the sub-datasets induces a distinct bias over the trained model. SDAT consists of several traversals, where each traversal means, a complete training cycle over all sub-datasets and the entire dataset.\\    
We first outline the details of our suggested training method and explain how it aids in steering the convergence towards a less biased solution by exploiting the specific characteristics of the collected sub-datasets. Then we provide a detailed explanation of our process for gathering these sub-datasets and how we generated them to possess their distinct characteristics as mentioned above. 


\vspace{-2.5mm}
\subsection{Training method}
\label{sub:cyclic_train}
Our training method as depicted in \Cref{fig:optimization} incorporates two primary mechanisms: 
\begin{enumerate}
    \item {\bf Dataset alternation:} The training process consists of alternating between different training phases. We alternate between a collected sub-dataset and the entire dataset. This alternation takes place every predetermined number of iterations. At each cycle, a different sub-dataset is used.
    \item {\bf Solution selection process:} During each training phase we monitor every iteration the average of all sub-datasets' validation loss along with the corresponding model's weights:
                   \begin{equation}
        \label{eq:Vt_calc} 
    \ \Bar{V}_t = \frac{1}{N}\sum_{i=1}^{N} \text{$V_{w_t,c_i}$} 
    \end{equation} 
            where $t$ is the iteration number (training step), $V_{w_t,c_i}$ is the average validation value of sub-dataset $c_i$, given model's weights $w_t$, and $N$ is the number of sub-datasets.
\begin{figure*}[ht!]
		\centering
        \begin{tabular}{ ccccc }
			\includegraphics[width=0.14\textwidth]{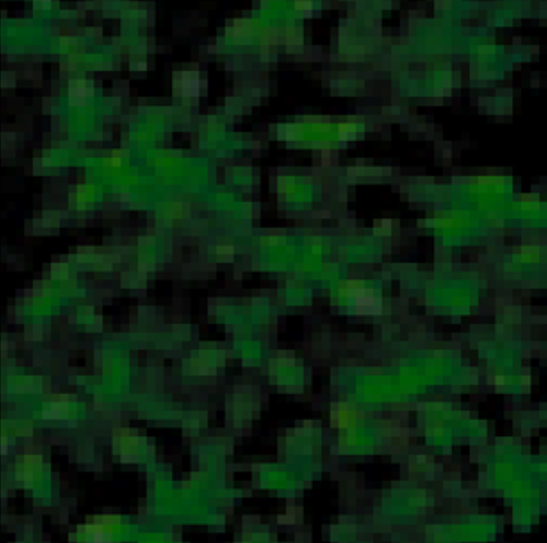} &
            \includegraphics[width=0.14\textwidth]{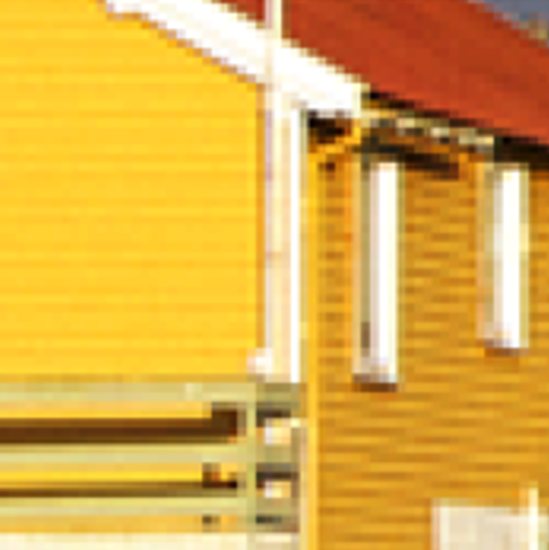} &
            \includegraphics[width=0.152\textwidth]{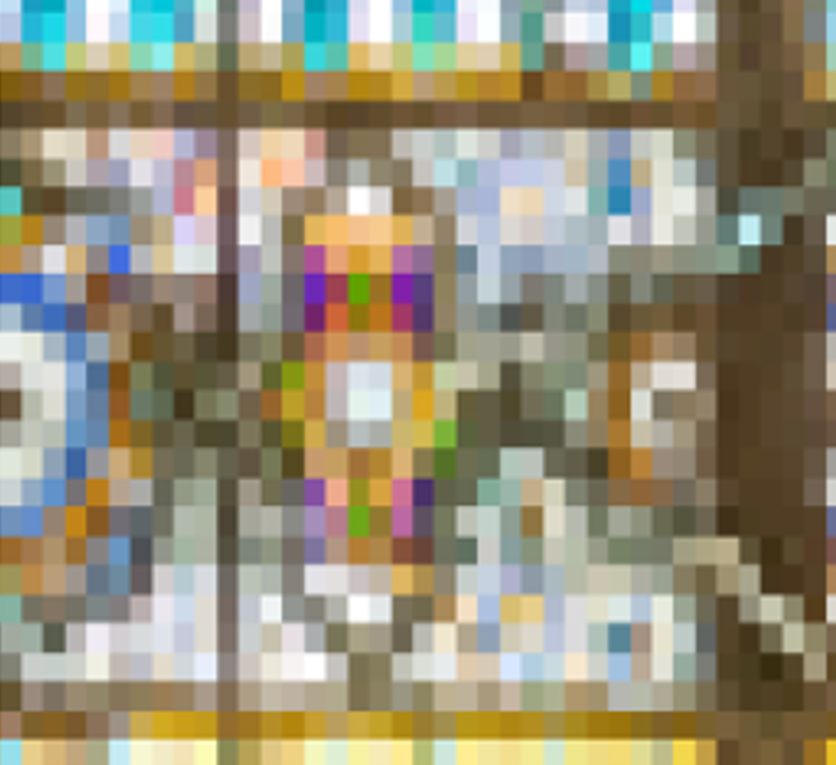} &
            \includegraphics[width=0.14\textwidth]{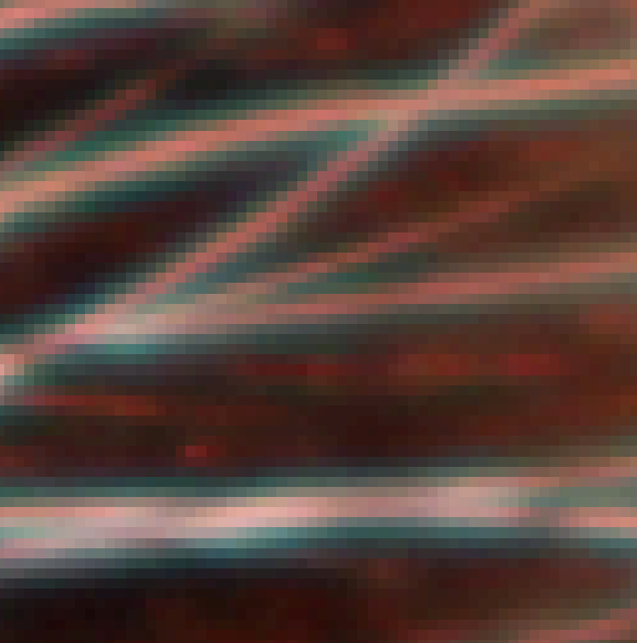} & 
			\includegraphics[width=0.141\textwidth]{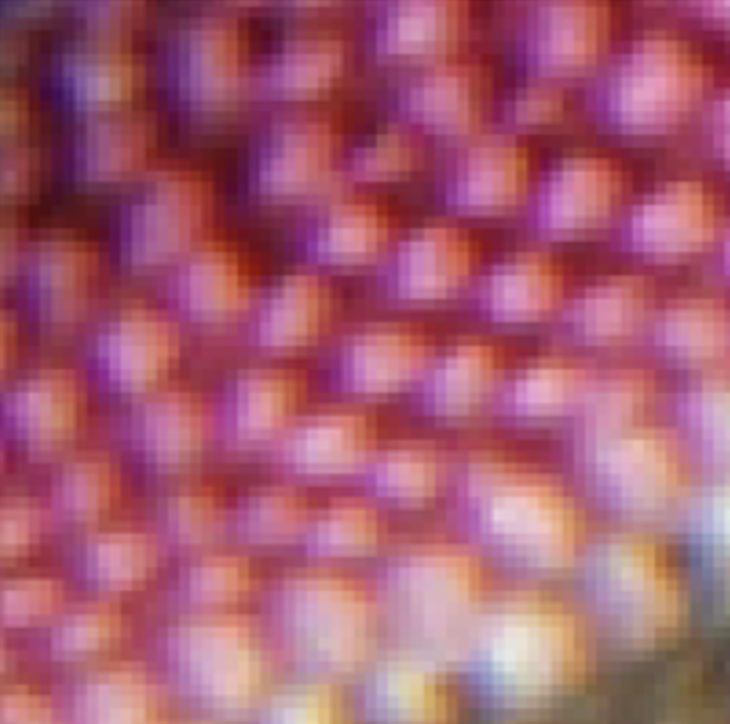} \\
            \includegraphics[width=0.14\textwidth]{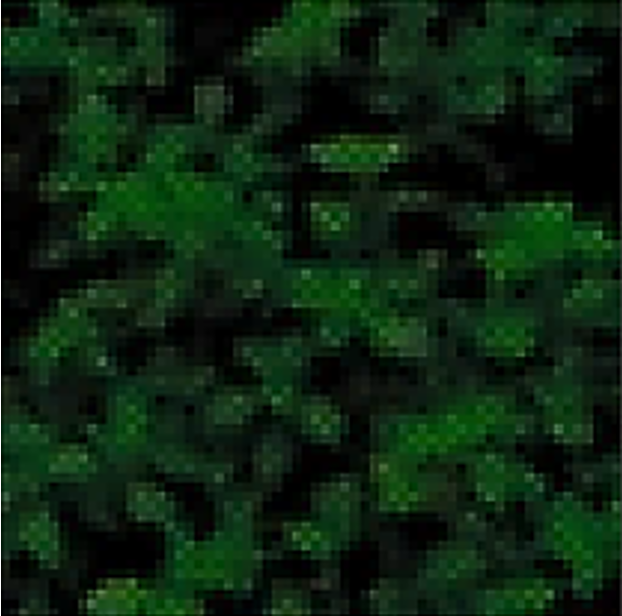}&
            \includegraphics[width=0.14\textwidth]{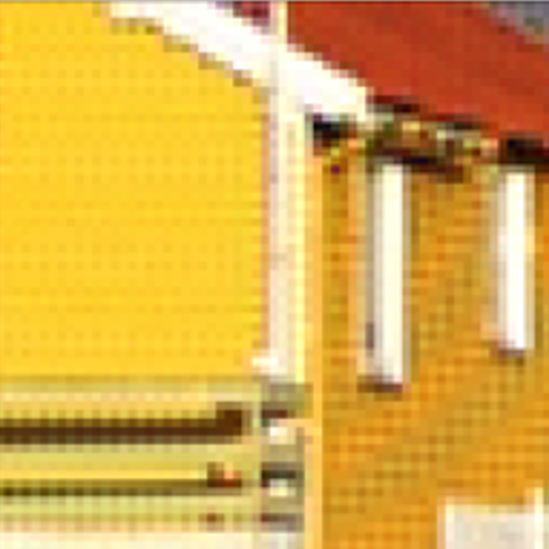}&
            \includegraphics[width=0.152\textwidth]{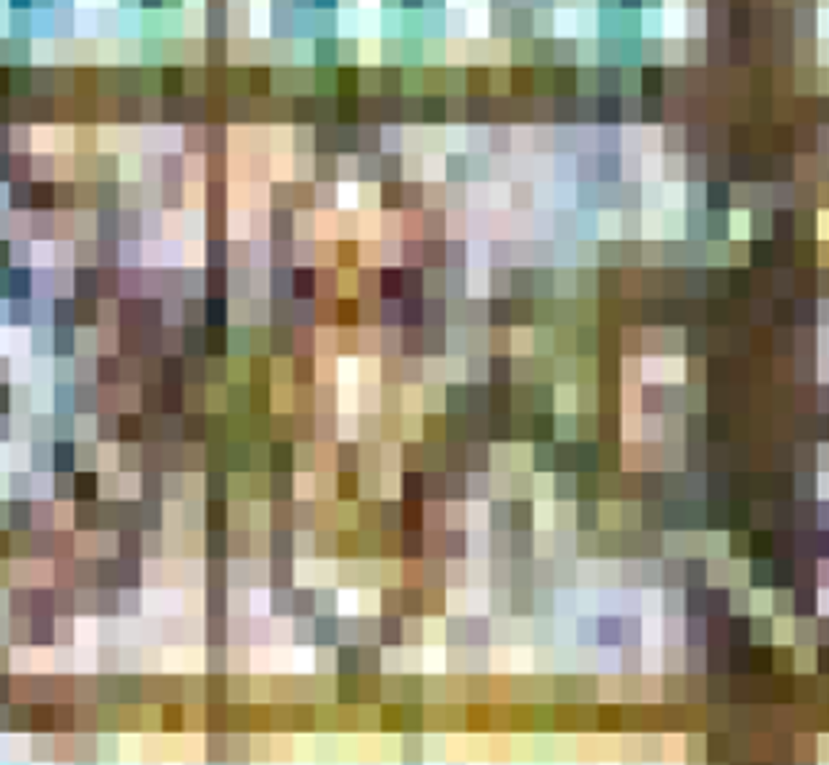} &
            \includegraphics[width=0.14\textwidth]{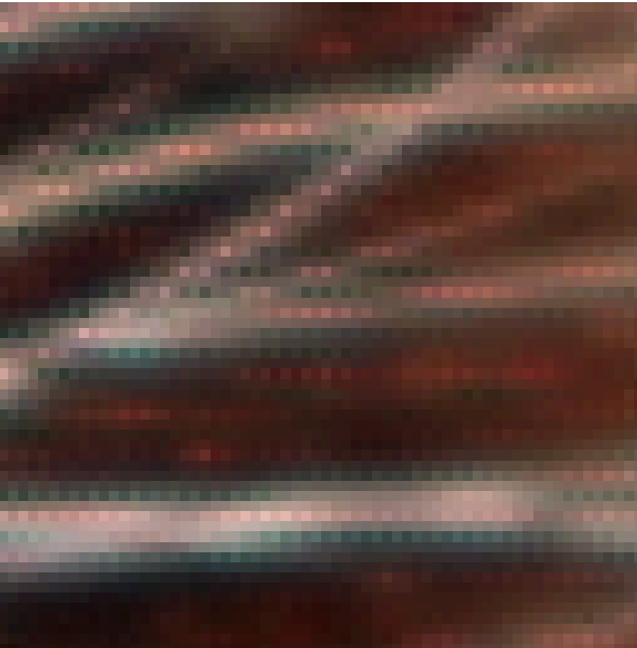}&
            \includegraphics[width=0.141\textwidth]{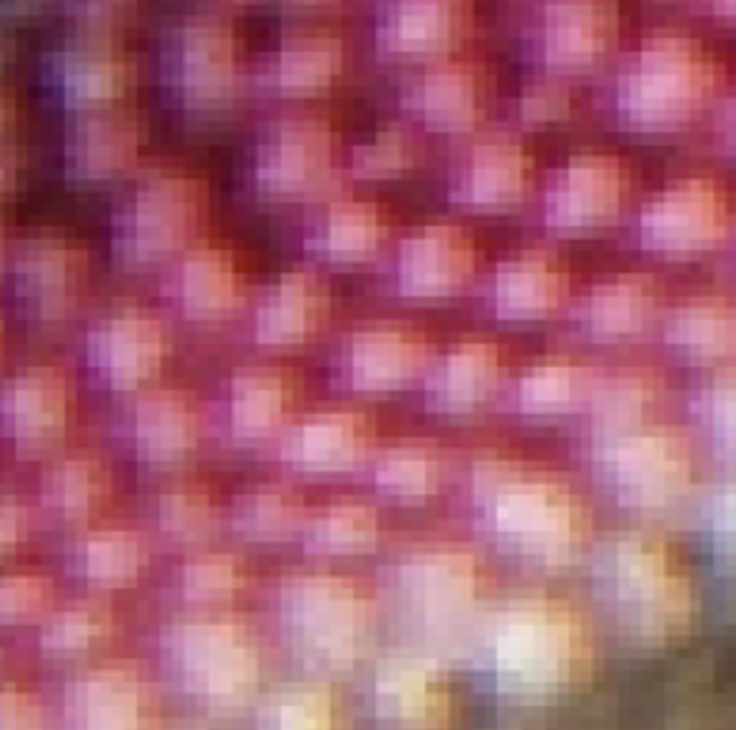} \\
            (a) Grid &
            (b) Zipper  &
            (c) $L_1$ &
            (d) Edge &  
            (e) Perceptual \\
\end{tabular}
		\caption{Examples of sub-datasets patches that are found via the different metrics. Each metric focuses on different aspects of image demosaicing. The top row depicts the ground truth patches, and the bottom is the prediction of the network after vanilla training. (a) A patch with grid artifacts, (b) a patch with zipper artifact, (c) a patch with high $L1$ distance, (d)  a patch with high edge distance, and (e) a patch with high perceptual distance.}
		\label{fig:mined_metrics}

\end{figure*}

    The chosen model's weights for the next alteration are the $w_t$ obtained the lowest $\Bar{V}_t$ during the training phase. See \cref{fig:sub_cat_train} for a more detailed example. 
     We observed that a model consistently achieving low validation loss across all sub-datasets is more likely 
    to converge towards a non-biased solution.
\end{enumerate}
\vspace{-2mm}
\noindent These two key factors combined help in guiding the convergence towards a less biased solution. Training over the entire dataset limits the exploration of the solution space due to the inherent bias of the dataset. By training over the sub-datasets, we are able to strongly steer the convergence towards a wider range of solutions, as each sub-dataset induces a different bias on the convergence.\\
Moreover, we maintain the usage of the entire dataset, as it contains the overall statistics of the task and every second alternation we train over the entire dataset. We are able to achieve a solution that does not converge to a biased local minimum due to the selection process, as it only propagates to the following training phase the least biased solution achieved during the current training phase.
 This way we are able to enjoy both worlds - we can still use the entire dataset for training, without being limited only to the solutions governed by the inherent dataset bias.\\ 
{\bf Sub-datasets convergence rate:} 
When training on the entire dataset, simple functions are converged first and complex functions are converged later as demonstrated in prior research \cite{rahaman2019spectral,basri2020frequency}. Our models also exhibited a similar trend. The initial convergence was governed by the entire dataset bias (in our case due to smooth patches in natural images), hindering further convergence toward more complex elements (high frequency, edges and more). In every training phase over a specific sub-dataset, the accumulation  of gradients is till the last non-biased update as in the following:  
    \begin{equation}
    \begin{aligned}
        \Gamma_{c_i} &= \sum_{t = t_{ci}}^{\hat{t}_{ci}} \frac{d}{dw_t} \mathcal{L}_{w_t, c_i} \\
        \hat{t}_{ci} &= \argmin_t \Bar{V}_t
    \end{aligned}
    \label{eq:GradsCalc}
\end{equation}
        where $\Gamma_{c_i}$ denotes the accumulated gradients obtained during a training session of a specific sub-dataset, $t_{ci}$ and $\hat{t}_{ci}$ denotes the training starting and ending time index respectively, using sub-dataset $c_i$ and  $\mathcal{L}_{w_t, c_i}$ denotes the training loss function given the model's weights $w_t$ and training over sub-dataset $c_i$.\\  After each traversal over all sub-datasets, the model's weights are composed based on the accumulated custom gradients from each sub-dataset:
        \begin{equation} 
        \label{eq:sub_cat}  
    w_{traversal} = w_{0} +\sum_{i=1}^{i=N} \Gamma_{c_i} +  \Gamma_{c_0}
    \end{equation}
        where $w_{0}$ is the weights of the model at the beginning of the traversal, $\Gamma_{c_0}$ is the accumulated gradients obtained along the entire dataset training phases. 

\noindent As demonstrated in \cite{rahaman2019spectral,basri2020frequency} and observed empirically in our experiments, controlling the convergence rate over different elements in the dataset leads to different solutions. Our training mechanism adapts the convergence rate of each sub-dataset as in \cref{eq:sub_cat}, in order to obtain the least biased solution.

{\bf Learning rate scheduler:}
To ensure stability during training, every dataset alternation we use a learning rate (LR) scheduler that starts with a low value and gradually increases to a higher value of LR.
Shifting between datasets significantly impacts the gradients at the current weight space position, altering the structure of the loss weight landscape and requiring readjustment of the LR. Only the weights that achieved the lowest validation loss across all sub-datasets in each training session are carried over to the next alternation as in \cref{eq:Vt_calc,eq:GradsCalc}, filtering out unstable solutions that might arise from unsuitable LR.

\begin{figure*}[ht!]
    \begin{center}
        \includegraphics[height=8.2cm, width=0.92\linewidth]{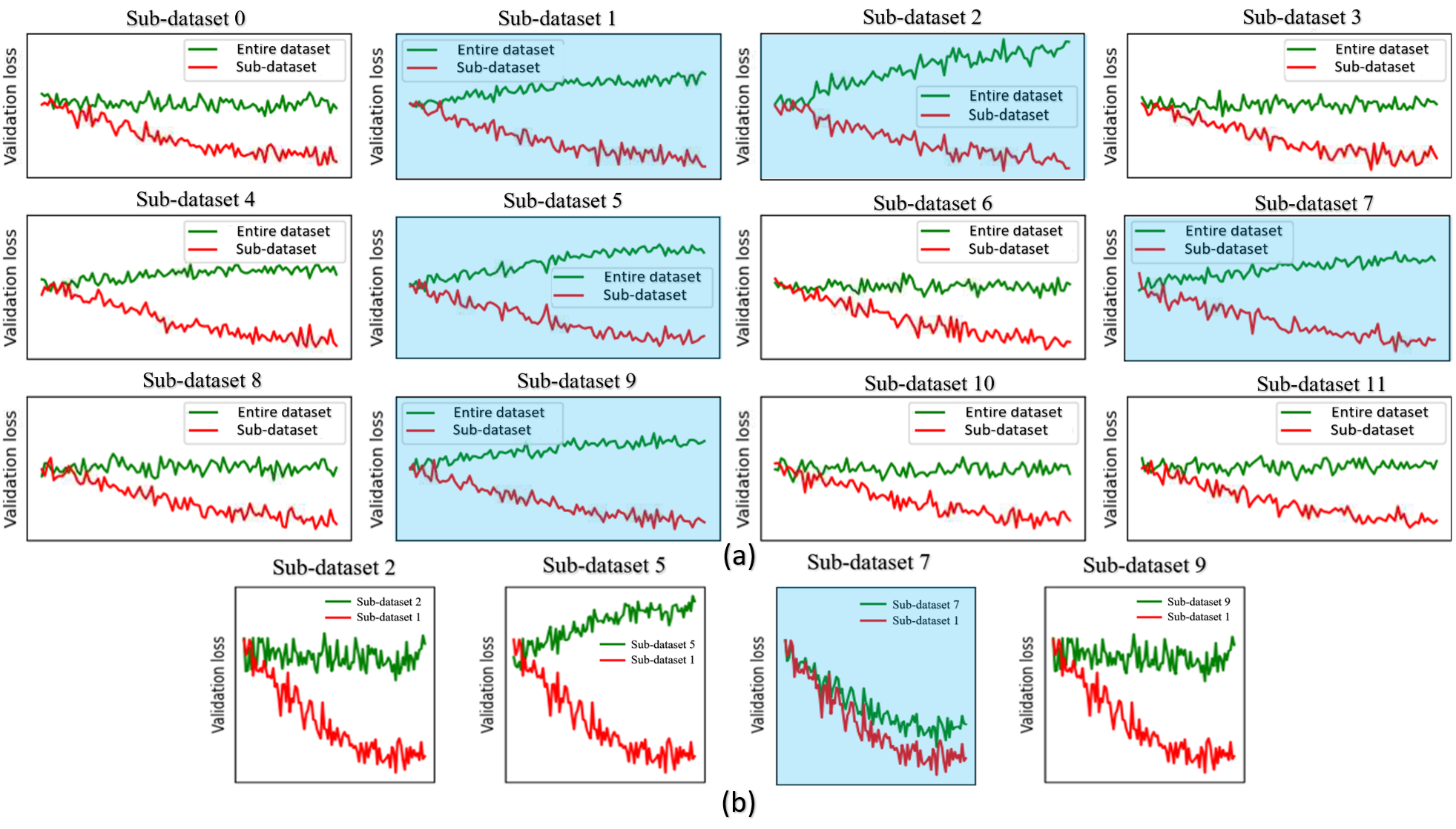}
    \end{center}         
        \caption{ Our two step elimination process for selecting the sub-datasets. (a) shows the first step, where each graph represents a training session over a sub-dataset (the horizontal axis shows the number of epochs, and the vertical axis shows the validation error). The green line represents the validation error of the entire dataset and the red line represents the validation error of the trained sub-dataset. Validation errors of sub-datasets with strong negative correlation convergence compared to the entire dataset validation error, are selected and highlighted in light blue. (b) depicts an example of the second step. The validation error of each sub-dataset (from those chosen in (a)) is compared with the validation error of the other sub-datasets. The objective is to merge the sub-datasets that demonstrate positive correlation convergence, resulting in a similar bias being induced by the trained model. In the example, we demonstrate the comparison of sub-dataset 1's validation (red line) with that of other chosen sub-datasets (green line). The sub-datasets chosen to be merged with other sub-datasets are highlighted in light blue.}
	\label{fig:UnCorr}   

\end{figure*}

\subsection{Sub-datasets identification and creation}  
We now describe the collection of sub-datasets from the entire dataset, each inducing a different bias on the trained model. This phase consists of the following:    
\begin{enumerate}
\label{sub_sec:idn_sub_cat2}
    \item {\bf Generation of sub-datasets}: To establish a starting point for data categorization, we performed a training using the entire dataset to obtain a base model. We used a combination of existing metrics, such as $L_1$, perceptual \cite{ledig2017photo}, and additional custom metrics, \cref{eq:HZ,eq:HG,eq:edge}, to detect a wide variety of artifact types created by the base model's outputs.
     we denote the initial predicted patch $\hat{P}_{init}$ and the patch in the ground-truth RGB image as $P$, both of size $H_p \times W_p \times 3$.
     \begin{enumerate}

    \item \textit{Zipper metric} - this metric $d_{zip}$ is used to find zipper artifacts near edges in the image. First, we calculate a mask $M_{NE}$ to identify non-edge areas as follows
    \begin{equation}
        \label{eq:MNE}
        M_{NE} := | C( \hat{P}_{init} ) - C\left( P \right) | < \varepsilon_1,
    \end{equation}
    where $C$ is some kind of edge detector, \eg Canny, and $\varepsilon_1$ is a predefined threshold.
    Then we are able to calculate
    \begin{equation}
        \label{eq:HZ}
        d_{zip} := \| | \, | \nabla \hat{P}_{init} | - \left| \nabla P \right| |  \odot M_{NE} \|_1,
    \end{equation}
    
    \item \textit{Grid metric} - this metric $d_{grid}$ is used to find grid-like artifacts in flat areas as follows
    \begin{multline}
        \label{eq:HG}
        d_{grid} := \| 2 \cdot | \nabla \hat{P}_{init} | \, \odot \\
        ( \sigma ( \frac{\alpha}{\left| \nabla P \right|}  ) -0.5 ) \odot ( | \nabla \hat{P}_{init} | > \varepsilon_2 ) \|_1,
    \end{multline}
    where $\sigma$ is the Sigmoid function, $\alpha > 0$ is a parameter that scales the values of the Sigmoid, and $\varepsilon_2$ is another predefined threshold.

        \item \textit{Edge distance} - This metric $d_{edge}$ is used to find distortions around edges
    \begin{equation}
	\label{eq:edge}
	    d_{edge} := \| C( \hat{P}_{init} ) - C\left( P \right) \|_1.
	\end{equation}
    \end{enumerate}

    We note that by using different thresholds in \cref{eq:MNE,eq:HG} we create a diverse pool of sub-datasets within the data. Examples of some of the identified patch categories are shown in \Cref{fig:mined_metrics}.
      
    \item {\bf Selection and elimination of sub-datasets}: From the collected pool of sub-datasets we keep the sub-datasets, that training on them induces a bias that hinders the generalization of the model on the entire dataset. These sub-datasets induce a different bias than the entire dataset and eventually help converge to a better solution. This procedure is illustrated in \Cref{fig:UnCorr} and is composed of the following two steps:
\begin{enumerate} 
\label{sub_sec:idn_sub_cat}
    \item {\bf Searching for inverse-correlation:} We conduct brief training sessions overall sub-datasets generated in the previous stage. Each session trains over a single sub-dataset and starts from the model that was trained over the entire dataset, as described in the previous stage. Throughout each session, we evaluate the validation loss over each of the different sub-datasets, including the entire dataset. We search for a negative correlation, in the sense of Spearman's rank correlation coefficient, between the validation loss on the trained sub-dataset and the validation loss on the entire dataset. If indeed there is a negative correlation, we identify a sub-dataset that the model struggles to generalize using the bias induced by the entire dataset. This means the model must discard its data-induced bias to achieve convergence for that specific sub-dataset.
    
    \item {\bf Merging sub-datasets}: When we detect sub-datasets with a positive correlation response between their validation losses (in the same sense as in step (a)), it means that training
    over one of the sub-datasets does not harm the model’s generalization over the other sub-dataset. Therefore, these sub-datasets are merged into a single sub-dataset.
          
\end{enumerate}        
\end{enumerate}      

In summary, we identify certain sub-datasets that are susceptible to bias, where each of the final generated sub-datasets is likely to induce a unique bias over the trained model. Since a model that has converged toward a biased solution is likely to produce a high validation loss (see \cref{eq:Vt_calc}) over these sub-datasets, it is used as a metric to estimate a biased solution.

\section{Experiments}
\label{sec:exp}
In order to evaluate our suggested training method we divided our experiments into two architecture groups that might be affected differently: (1) Low-capacity models and (2) High-capacity models.
Over each architecture, we conducted two training types: the first is a training over the entire dataset, which we refer to as EDT (Entire Dataset Training) and the second is our proposed SDAT method.\\
For both types of training settings, we used crops of size 64 $\times$ 64 as the input and a batch size of 16. We also used data augmentations, such as a random flip and transpose over both image axes during training. All models were trained using the $L_{1}$ norm between the ground truth and estimated RGB images as the loss function. \\
\textbf{EDT method settings:} we used the DIV2K \cite{agustsson2017ntire} dataset for training. 
We used the Adam optimizer \cite{kingma2014adam} with learning rate $5 \times 10^{-4}$,  $\beta_{1} = 0.9 $, $ \beta_{2} = 0.999 $, and $ \epsilon = 1 \times 10^{-8}$. The models were trained for a total of 2M iterations. \\
\textbf{SDAT method settings:} In addition to using the DIV2K dataset as a whole, we employed the process described in \Cref{sub_sec:idn_sub_cat2}, and chose 5 sub-datasets obtained from the DIV2K dataset. These sub-datasets demonstrated the strongest negative correlation with the entire dataset. We used the same optimizer parameters as in the standard training, except for the suggested modification to the LR scheduler, as explained in \Cref{sub:cyclic_train}. We trained for 20 traversals (each traversal means, a complete training cycle over all sub-datasets and the entire dataset). Each traversal consists of 100K iterations.

It should be noted that SDAT takes more time to converge compared to EDT because of its more complex data setup. To ensure a fair comparison, we've chosen to run the same number of training cycles for both methods.

\subsection{Low-capacity models}
\label{sub:low_capacity}
We trained three CNN architectures with 9.5K, 16K, and 84K parameters. Our CNN has a similar structure to the overall structure of DemosaicNet suggested by Gharbi \etal in \cite{gharbi2016deep}, with slight differences. We replaced each of the two convolution layers with 3 Inverted Linear bottlenecks (ILB) \cite{sandler2018mobilenetv2}. We obtain different architecture sizes by adjusting the expansion size parameters of the ILB. In addition to comparing the results between EDT and our suggested method, we also compared our results (in terms of PSNR) with recent studies that focus on achieving high performance for image demosaicing using low capacity models \cite{ramakrishnan2019deep,wang2021compact}. As indicated in \Cref{tb:low_capacity}, our training method presents superior results across all benchmarks compared to the EDT method. Additionally, while our trained CNNs have the lowest number of parameters in each sub-range of parameters, the networks outperform all other methods across all benchmarks.

\begin{table}[!t]\footnotesize
    \centering
    \begin{tabular}{|c|c|c|c|c|c|}
    \hline
        Parameter & \multirow{2}{*}{Method} & \multirow{2}{*}{Params.} & \multirow{2}{*}{Kodak } & \multirow{2}{*}{MCM} & \multirow{2}{*}{Urban100} \\ 
        range & ~ & ~ &  &  &  \\ \hline \hline
        \multirow{3}{*}{9K-12K} & Ark Lab \cite{ramakrishnan2019deep} & 10K & 40.4 & 36.9 & - \\   
        ~ & Wang \cite{wang2021compact} & 11.7K & 40.8 & 36.75 & 36.4 \\ 
        ~ & CNN EDT & 9.5K & 38.3 &35.4 & 35.44 \\ 
        ~ & \bf{CNN SDAT} & 9.5K & \textbf{41.39} & \textbf{37.02} & \textbf{36.79} \\
        \hline 
        \multirow{2}{*}{16K-20K} & Ark Lab & 20K & 40.9 & 37.6 & - \\    
         ~ & CNN EDT & 16K & 39.31 & 35.81 & 36.2 \\
        ~ & \bf{CNN SDAT} & 16K & \textbf{42.05} & \textbf{37.80} & \textbf{37.52} \\ \hline
        \multirow{3}{*}{80K-200K} & Ark Lab & 200K & 41.2 & 38.2 & - \\ 
        ~ & Wang & 183K & 41.72 & 37.91 & 37.7 \\  
         ~ & CNN EDT & 84K & 40.25 & 37.44 & 37.04 \\ 
        ~ & \bf{CNN SDAT} & 84K & \textbf{42.38} & \textbf{38.52} & \textbf{38.01} \\ \hline
    \end{tabular}
    \caption{PSNR results for image demosaicing for compact networks. The best results are highlighted in bold.}
\label{tb:low_capacity}
\end{table}

\subsection{ High-capacity models}
\label{sub:transformers_arc}
We evaluated our SDAT method in comparison to the EDT method over three leading transformer architectures. 
First, we implemented RSTCANet \cite{xing2022residual}, which is based on the SwinIR \cite{liang2021swinir} architecture. We applied both methods over two RSTCANet variants, RSTCANet-B and RSTCANet-S with 0.9M and 3.1M parameters respectively. Additionally, we trained a variant of the GRL-S \cite{li2023efficient} architecture with 3.1M parameters. 

As can be seen in \Cref{table:our_vs_standard}, we evaluated four popular benchmarks that represent natural images, with additional two benchmarks HDR-VDP and Moir\'e initially presented in \cite{gharbi2016deep}, and are known as a collection of hard patches. Our training method overall architectures produced superior results across all benchmarks in comparison to the EDT method described above. 
It is worth mentioning, that while the EDT method yielded comparable or even superior results to the published results across all RSTCANet variants, it could not reconstruct the reported results for the GRL. However, as can be seen in \Cref{tb:SOTA} by using the SDAT method to train the GRL architecture, we achieved SOTA over three popular benchmarks and second best in the fourth.

\begin{table*}[!ht]\normalsize  
\centering
    \begin{tabular}{|l|c|c|c|c|c|c|}
    \hline
        Method &MCM &  Kodak &CBSD68  &  Urban100& HDR-VDP& Moir\'e\\  \hline
        RSTCANet-B-EDT & 38.83 & 42.22 &  41.83 & 38.45 & 33.6& 36 \\ \hline
        RSTCANet-B-SDAT &  \textbf{39.72} &\textbf{43.22} & \textbf{42.58} & \textbf{39.88} & \textbf{34.52}& \textbf{37.81} \\\hline
        \hline
        RSTCANet-S-EDT & 39.52 & 42.71 & 42.44 &39.72  & 34.34& 37.53 \\ \hline
        RSTCANet-S-SDAT & \textbf{39.75} & \textbf{43.31} & \textbf{42.82} & \textbf{40.14} & \textbf{34.72}&  \textbf{37.92}\\\hline
        \hline  
        GRL-EDT & 39.7 & 43.14& 42.95 &40.05 & 34.7 & 38.1\\ \hline
       GRL-SDAT &  \textbf{40.11} &\textbf{43.65} & \textbf{43.31} & \textbf{40.67}& \textbf{34.95} & \textbf{38.3} \\\hline
    \end{tabular} 
        \vspace{2mm}

    \caption{PSNR results for image demosaicing for leading transformer architectures. We compare standard training over the entire dataset (EDT) with our suggested training approach (SDAT). The best results for each architecture are in bold.}  
\vspace{0.35cm}

\label{table:our_vs_standard}
\end{table*}
\vspace{2mm}
 \begin{table*}[!h]  
\center    
\begin{center}
\small
\begin{tabular}{|l|c|c|c|c|c|c|c|c|c|c|c|c|c|c|c|c|c|}
\hline
\multirow{2}{*}{Method} &  \multirow{2}{*}{Params.} & \multicolumn{2}{c|}{MCM} &  \multicolumn{2}{c|}{Kodak} &  \multicolumn{2}{c|}{CBSD68} &  \multicolumn{2}{c|}{Urban100}   
\\
\cline{3-10}
 & & PSNR & SSIM & PSNR & SSIM & PSNR & SSIM & PSNR & SSIM 
\\
\hline
\hline


DRUNet\cite{zhang2021plug}
& 11M & 39.40 & \underline{0.991} & 42.30 & 0.994 & 42.33 & 0.995 & 39.22 & 0.990
\\

RNAN \cite{zhang2019residual}
& 9M & 39.71 & 0.972 & 43.09 & 0.990 & 42.50 & 0.992 & 39.75 & 0.984
\\

JDD \cite{xing2021end}
& 3.5M & 38.85 & 0.990 & 42.23 & 0.994 & - & - & 38.34 & 0.989
\\
PANet \cite{mei2023pyramid}
& 6M & 40.00 & 0.973 & 43.29 & 0.990 & 42.86 & 0.989 & 40.50 & 0.985
\\

RSTCANet-B \cite{xing2022residual} 
& 0.9M & 38.89 & 0.990 & 42.11 & 0.994 & 41.74 & 0.9954 & 38.52 & 0.990
\\

RSTCANet-S \cite{xing2022residual} 
& 3.1M & 39.58 & 0.991 & 42.61 & 0.995 & 42.36 & 0.995 & 39.69 & 0.992
\\

RSTCANet-L \cite{xing2022residual} 
& 7.1M & 39.91 & \textbf{0.991} & 42.74 & 0.995 & 42.47 & 0.996 & 40.07 & \underline{0.993}  
\\
GRL \cite{li2023efficient} 
& 3.1M & \underline{40.06} & 0.988 & \underline{43.34} & \underline{0.995} & \textbf{43.89} & \textbf{0.997} & \underline{40.52} & 0.992  
\\
\hline   
RSTCANet-B \bf{SDAT}
& 0.9M & 39.72 & 0.98 & 43.22 & 0.995 & 42.58 & 0.996 & 39.88 & 0.991
\\

RSTCANet-S \bf{SDAT} 
& 3.1M & 39.75 & 0.990 & 43.31 & 0.995 & 42.82 & 0.996 & 40.14 & \textbf{0.993 } 
\\ 
GRL \bf{SDAT}
& 3.1M & \textbf{40.11} & 0.988 & \textbf{43.65} & \textbf{0.996} & \underline{43.31} & \underline{0.996} & \textbf{40.67} & 0.992 
\\ 
\hline  
\end{tabular}
\end{center}
    \vspace{2mm}

\caption{ Benchmark results (PSNR and SSIM) for image demosaicing. The best results are in bold, and the second best are underlined. Except for the results obtained by our SDAT  method, the rest of the results in this table were obtained by the official code provided by the authors or reported by them in the original papers.}
\label{tb:SOTA}
\vspace{0.2cm}
\end{table*}
\vspace{0.2cm}


\begin{table*}[!th]
\center  
\small
\begin{tabular}{|c|cc|cc|cc|cc|cc|}
\hline
\multirow{3}{*}{Params.} & \multicolumn{2}{c|}{\multirow{2}{*}{EDT}} & \multicolumn{2}{c|}{\multirow{2}{*}{Mined training}} & \multicolumn{2}{c|}{\multirow{2}{*}{\begin{tabular}[c]{@{}c@{}} Uniform \\ distribution training\end{tabular}}} & \multicolumn{2}{c|}{\multirow{2}{*}{\begin{tabular}[c]{@{}c@{}}SDAT\\  without EDT phase\end{tabular}}} & \multicolumn{2}{c|}{\multirow{2}{*}{SDAT}} \\
 & \multicolumn{2}{c|}{} & \multicolumn{2}{c|}{} & \multicolumn{2}{c|}{} & \multicolumn{2}{c|}{} & \multicolumn{2}{c|}{} \\ \cline{2-11} 
 & \multicolumn{1}{c|}{Kodak} & MCM & \multicolumn{1}{c|}{Kodak} & MCM & \multicolumn{1}{c|}{Kodak} & MCM & \multicolumn{1}{c|}{Kodak} & MCM & \multicolumn{1}{c|}{Kodak} &MCM \\ \hline
16K & \multicolumn{1}{c|}{39.20} & 35.90 & \multicolumn{1}{c|}{ \underline{40.20} } & 36.40 & \multicolumn{1}{c|}{40.10} & \underline{37.50} & \multicolumn{1}{c|}{39.70} & 37.50 & \multicolumn{1}{c|}{ \textbf{42.05} } & \textbf{37.80} \\ \hline
176K & \multicolumn{1}{c|}{40.55} & 37.20 & \multicolumn{1}{c|}{40.90} & 37.50 & \multicolumn{1}{c|}{ \underline{41.10} } & \underline{38.20} & \multicolumn{1}{c|}{41.00} & 38.00 & \multicolumn{1}{c|}{ \textbf{42.41} } & \textbf{38.67} \\ \hline
\end{tabular}
    \vspace{2mm}
\caption{ PSNR results of various training methods compared to ours over the Kodak and McMaster datasets. We evaluated each training method using two architectures consisting of 16K and 176K parameters. The best results are in bold, and the second best are underlined.}
\label{tb:mined_data}
\vspace{0.1cm}
\end{table*}


\subsection{Ablation Study}
{\bf Comparing different training methods:} 
To verify the contribution of our suggested training method we conducted the following ablation study to compare our method's effectiveness versus other training methods: (1) standard training over the entire dataset, (2) standard training over hard mined examples only (similar to \cite{gharbi2016deep}), (3) standard training over a uniform distribution, i.e. we compose a new dataset where the probability to sample a training example from the entire dataset and from the sub-datasets is equal, and (4) performing our method, but discarding the alternation to train on the entire dataset. The training procedure for the standard training methods (1)-(3) is the same as described for EDT settings above. As indicated in \Cref{tb:mined_data}, our method surpasses all other training regimes, by a considerable margin. 
Comparing to standard training, we conducted a wider experiment covering a wide range of model sizes using the same training procedure described above.
The results are shown in \Cref{fig:intro}. It is easy to see how we are able to improve the networks' performance between 1-3.5dB depending on the initial model size (the smaller the network, the higher the performance boost) and the evaluation dataset. \\
{\bf Comparing different numbers of sub-datasets:}
As outlined in \Cref{sub:cyclic_train}, our approach involves switching between a collection of sub-datasets and the entire dataset. In this experiment, we assess the effectiveness of our training method by randomly selecting one and two sub-datasets. The results, as shown in \Cref{tb:num_categories}, demonstrate that the number of sub-datasets included in the alternation process has a significant impact on the resulting performance. This demonstrates that being able to identify additional types of bias in the original dataset can further improve the network's overall generalization.

\vspace{-1mm}
\begin{table}[!t] \footnotesize
    \centering
    \begin{tabular}{|l|c|c|c|c|c|c|c|c|}
    \hline
        \multirow{2}{*}{Params.}   & \multicolumn{2}{c|}{1 sub-dataset} & \multicolumn{2}{c|}{2 sub-datasets} & \multicolumn{2}{c|}{5 sub-datasets} \\ 
        \cline{2-7}
        ~   & Kodak  & MCM  & Kodak  & MCM  & Kodak  & MCM  \\ \hline
        16K & 39.22  & 36.21 & 40.50 & 36.54 & \bf{42.05} & \bf{37.80} \\ \hline
        176K & 40.50  & 37.25 & 41.25 & 37.77 & \bf{42.41} & \bf{38.67} \\ \hline
    \end{tabular}
    \vspace{2mm}
            \caption{ PSNR results over Kodak and MCM datasets using a different number of sub-datasets for training. The best results are in bold.}
\label{tb:num_categories}      
\end{table}
\section{Conclusion} 
In this paper, we addressed the challenge of dataset inductive bias for the image demosaic task and proposed SDAT, a novel training method to improve the performance of a model. The method involves two main steps: defining and identifying sub-datasets beneficial to model convergence and an alternating training scheme. 
SDAT demonstrated an improved performance for image demosaicing over standard training methods and achieved state-of-the-art results on three highly popular benchmarks. Our suggestion also effectively utilized the model's capacity, surpassing recent relevant works on low-capacity demosaicing models across all benchmarks using fewer parameters. This result is crucial for edge devices. The method's success also demonstrates the importance of considering the dataset structure in optimizing a model's training process. Our findings suggest that our proposed method can be applied to other image restoration tasks and can be used as a trigger for further research in this field.

\newpage
\bibliographystyle{IEEEtran} 
\bibliography{main}

\end{document}